\documentclass[letter, 10pt,twocolumn,comsoc]{IEEEtran}

\IEEEoverridecommandlockouts
\usepackage[pdftex]{graphicx}
 \usepackage{lipsum,graphicx,subcaption}
\usepackage{float}
\usepackage{cite}
\usepackage{amsmath,amssymb,amsfonts}
\usepackage{algorithmic}
\usepackage{textcomp}
\usepackage{algorithmic,algorithm}
\usepackage{url}
\usepackage[11pt]{moresize}
\def\BibTeX{{\rm B\kern-.05em{\sc i\kern-.025em b}\kern-.08em
    T\kern-.1667em\lower.7ex\hbox{E}\kern-.125emX}}
\begin{document}

\title{Self-Organizing mmWave Networks : A Power Allocation Scheme Based on Machine Learning}

\author{
    \IEEEauthorblockN{Roohollah Amiri\IEEEauthorrefmark{1}, Hani Mehrpouyan\IEEEauthorrefmark{1}}\\
    \IEEEauthorblockA{\IEEEauthorrefmark{1}\textit{\small{School of Electrical and Computer Engineering, Boise State University,\\
    \{roohollahamiri,hanimehrpouyan\}@boisestate.edu}}}
}

\maketitle

\begin{abstract}
Millimeter-wave (mmWave) communication is anticipated to provide significant throughout gains in urban scenarios. To this end, network densification is a necessity to meet the high traffic volume generated by smart phones, tablets, and sensory devices while overcoming large pathloss and high blockages at mmWaves frequencies. These denser networks are created with users deploying small mmWave base stations (BSs) in a plug-and-play fashion. Although, this deployment method provides the required density, the amorphous deployment of BSs needs distributed management. 
To address this difficulty, we propose a self-organizing method to allocate power to mmWave BSs in an ultra dense network. The proposed method consists of two parts: clustering using fast local clustering and power allocation via Q-\textit{learning}. 
The important features of the proposed method are its scalability and self-organizing capabilities, which are both important features of 5G. Our simulations demonstrate that the introduced method, provides required quality of service (QoS) for all the users independent of the size of the network.
\end{abstract}

\vspace{-15pt}
\section{Introduction}
Millimeter-wave (mmWave) communication is one of the main technologies of the next generation of cellular networks (5G). The large bandwidth at mmWave frequency has the potential to enhance network throughput by tenfolds~\cite{art_rappaport}. However, large path loss and shadowing limit the performance of mmWave systems and need to be dealt with. One approach to overcome this problem is based on increasing the density of access points~\cite{art_dense0,art_dense1}. However, as the number of access points increases, the complexity of network management increases. Keeping this in mind, one of the features of future mmWave base stations (BSs) is self-deployment by users. In other words access points can be deployed in a plug-and-play fashion, and the network architecture may change frequently. Considering the above points, 5G needs self-organizing methods to configure, adapt, or heal itself when necessary. In this paper, a self-organizing algorithm is proposed to maximize the sum capacity in a dense mmWave network while providing users with their required quality of service (QoS). The algorithm consists of clustering, based on fast local clustering (FLOC), and distributed power allocation, via Q-\textit{learning}. Scalability and fast convergence of FLOC, adaptability and distributed nature of Q-\textit{learning}, makes their combination a suitable tool to achieve self-organization in a dense network.

\section{System Model}\label{sec_systemModel}
The System model considers a dense outdoor urban scenario as an important example of 5G, i.e., we consider the downlink of densely deployed mmWave BSs. 
To this end let us consider $N$ mmWave BSs that are distributed based on the homogeneous spatial Poisson point process (SPPP) with density $\lambda_{BS}$~\cite{art_poisson}. Each BS is associated to one user. BSs share a single frequency resource block (FRB) to support their associated users. We assume a time invariant channel model, i.e. slow fading. The channel vector between the BS $i$ and user $k$, can be written as follows
{
\begin{align}\label{eq_channel}
H_{i,k}=\left(L_{i,k}\right)^{-1} \times g_{i,k},
\end{align}
}
where $L_{i,k}$ and $g_{i,k}$ denote the path loss and the path gain between the BS $i$ and user $k$. The path loss between the BS $i$ and its associated user $i$, $L_{i,i}$, follows the free space propagation based on Frii's law~\cite{art_rappaport}. Here, we consider that the majority of interferers have non-line-of-sight (NLOS) paths\cite{art_Intf_2}. Hence, the path loss $L_{i,k}$ ($i\neq k$) can be written as~\cite{art_rappaport}
{
\begin{align}\label{eq_loss}
L_{i,k} [dB] = \beta_1 + 10 \beta_2 \log_{10}\left(d_{i,k}\right) + X_\zeta,
\end{align}
}
where $\beta_1$ and $\beta_2$ are factors used to achieve best fit to channel measurements, $d_{i,k}$ is the distance between the BS $i$ and the user $k$, $X_\zeta$ denotes the logarithmic shadowing factor, where $X_\zeta\thicksim \mathcal{N}\left(0,\zeta^2\right)$, and $\zeta^2$ denotes the lognormal shadowing variance.

The received signal in the downlink at the $k^{\text{th}}$ user includes the desired signal from its associated BS (BS $k$), interference from neighboring BSs, and also thermal noise. Hence, the signal-to-interference-noise-ratio (SINR) at the $k^{\text{th}}$ user is given by
{
\begin{align}\label{eq_sinr_ue}
\text{SINR}_{k}=\frac{P_k H_{k,k}}{ \sum_{i\in D_k,i\neq k}^{~} P_i H_{i,k} + \sigma^2},
\end{align}
}
where $P_k$ denotes the power transmitted by the $k^{\text{th}}$ BS, $D_k$ is the set of interfering BSs, and $\sigma^2$ denotes the variance of the additive white Gaussian noise. Accordingly, the normalized capacity at the $k^{\text{th}}$ user is given by
{
\begin{equation}\label{eq_c1}
C_{k} = \log_2 (1+\text{SINR}_{k}).
\end{equation}
}


\vspace{-15pt}
\section{Problem Formulation}\label{sec_problem}
The goal of the optimization problem is to find the best power distribution between mmWave BSs ($\bar{P}$) in order to maximize the sum capacity of the network, while supporting all users with their required QoS. The optimization problem ($\textbf{P}_1$) can be formulated as
{
\begin{subequations}\label{opt_1}
\begin{align}
& \underset{\bar{P}}{\text{maximize}}
& & \sum_{k=1}^N \log_2 (1+\text{SINR}_{k})  \label{a}\\
& \text{subject to}
& & P_k \leq P_{max}, \; k = 1, \ldots, N\label{b}\\
&&& \text{SINR}_{k} \geq q_k , \;k = 1,...,N.\label{c}
\end{align}
\end{subequations}
}
Here, the objective \eqref{a} is to maximize the sum capacity of the network while providing all users with their required QoS in \eqref{c}. The first constraint, \eqref{b}, refers to the power limitation of every BS. The term $q_k$ in \eqref{c} refers to the minimum required $\text{SINR}$ for the $k^{\text{th}}$ user.

Eq.~\eqref{a} contains the interference term in the denominator of $\text{SINR}$ term. In a dense network the interference term cannot be ignored~\cite{art_slmz}. Due to the presence of the interference term, the objective function~\eqref{a} is a non-concave function~\cite{art_cvx_1}.

The solution to $\textbf{P}_1$ should have certain features. First, it should be distributed due to no central authority in this network. Second, the range of mmWave BSs is limited, so each user will receive interference from the BSs in its neighborhood. Therefore, the solution should consider local clustering to reduce the computation overhead. Third feature is self-healing. The number of BSs in the network changes sporadically, which means the solution should be adaptive to new possible architectures. Considering the above, in this paper, we propose a method which contains two parts : a fast local clustering method to locally cluster the BSs, and in each cluster, BSs will choose their transmitting power based on Q-\textit{learning}~\cite{Watkins1992}. Q-\textit{learning} is model-free (adaptable) and gives the BSs the ability to learn from their environment by interacting with it (self-organization). 

\section{Cluster Based Distributed Power Allocation Using Q-Learning (CDP-Q)}\label{sec_solution}

In our proposed method, mmWave BSs are considered as the agents of Q-\textit{learning}, so the terms agent and mmWave BS are used interchangeably. CDP-Q is a distributed method in which multiple agents (mmWave BSs) find a sub-optimal policy (power allocation) to maximize the network capacity. CDP-Q consists of two parts: (1) clustering, and (2) power allocation. Clustering is based on a local clustering method, and power allocation is based on Q-\textit{learning}. In the following each part is detailed.

\vspace{-5pt}
\subsection{MmWave BSs Clustering}\label{sec_floc}


Since mmWave signals suffer from high pathloss and shadowing, only neighboring BSs that are close in distance interfere with each other. Consequently, we propose to use a clustering mechanism to divide BSs into clusters in which the interference of one cluster is negligible on other clusters' users.




In this paper, we propose to use Fast local clustering (FLOC)~\cite{art_floc} to divide mmWave BSs into clusters. FLOC is a distributed message-passing clustering method with $\mathcal{O}(1)$ complexity, which guarantees scalability, and produces non-overlapping clusters. Another feature of FLOC is local self-healing, which means re-clustering, due to addition of a new node or removing a node, does not propagate through all clusters. In order to apply FLOC in a mmWave network, the following concepts are defined:

\begin{itemize}
\item \textit{Cluster head (CH)}: The mmWave BS that is chosen as the head of the cluster. In our algorithm, there is no priority between a cluster head and other members of the cluster.
\item \textit{In-bound (IB), and out-band (OB) node}: In FLOC, a node is in-bound if it is a unit distance from a CH. A unit distance is a set value, which in this case is the range of mmWave links, i.e., $100$-$200~\text{m}$~\cite{art_rappaport}. Accordingly, we define in-bound as $100~\text{m}$, which is an indication of strong interference, and out-band as $200~\text{m}$, which indicates the edge of the cluster around a CH. Finally, if a node $j$ is in out-band distance of a cluster $i$, and not in an in-bound distance of any other clusters, then node $j$ will join the cluster $i$ as an OB node. 
\end{itemize}

\vspace{-15pt}

\subsection{Distributed Power Allocation Using Q-Learning}\label{sec_powerAlloc}
The output of Q-\textit{learning} is a decision policy (power allocation) which is represented as a function called Q-function. Here, the Q-function of agent $k$ is represented as a table called a Q-table ($Q_k$). The columns of a Q-table are the actions ($a^k$), and the rows are the state ($\mathbf{s}^k$) of the agent $k$. 

In multi-agent Q-\textit{learning}, agents can act independently or cooperatively. In the independent learning, each agent interacts with the environment without communicating with other agents. In fact, it considers the other agents as part of the environment. Independent learning has shown good performance in many applications~\cite{art_Panait}. In independent learning, since the environment is not stationary, oscillation and longer convergence time might happen for the agents, but due to no communication overhead between agents compared to cooperative learning, we choose independent learning. Motivated by this fact, the agents will select their actions according to~\cite{book_sutton}
{
\begin{align}\label{eq_IL_learning}
a_t^k =  \arg\max_{a} Q_k\left(\mathbf{s}_t^k,a\right),
\end{align}
}
in which, subscript $t$ denotes time step $t$ of Q-\textit{learning}. The CDP-Q algorithm is presented in Algorithm 1.
\begin{algorithm}[h]
\renewcommand\thealgorithm{}
\caption{1 The proposed CDP-Q algorithm}\label{alg_2}
\begin{algorithmic}[1]{
\STATE {Cluster formation based on Sec.~\ref{sec_floc}}
\FORALL{Clusters in Parallel}
\FORALL{Agents}
\STATE Initialize $Q(\mathbf{s}_t^k,a_t)$ arbitrarily
\STATE Initialize $\mathbf{s}_t^k$
\FORALL{episodes}
\STATE send $Q_k\left(\mathbf{s}_t^k,:\right)$ to other agents of the cluster
\STATE receive $Q_j\left(\mathbf{s}_t^j,:\right),j \in D_k,j\neq k$ 
\STATE Choose $a_t^k$ according to Eq.~\ref{eq_IL_learning}
\STATE Take action $a_t^k$, observe $R_t^k$
\STATE \footnotesize\begingroup
   $Q_k(\mathbf{s}_t^k, a_t^k) \leftarrow (1-\alpha)Q_k(\mathbf{s}_t^k, a_t^k) + \alpha(R_t^k + \gamma Q_k(\mathbf{s}_{t+1}^k, a_t^k))$
\endgroup
\normalsize
\ENDFOR
\ENDFOR
\ENDFOR
}
\end{algorithmic}
\addtocounter{algorithm}{-1}
\end{algorithm}

In the following the actions, states, and the reward function of the proposed Q-\textit{learning} method are defined.
\subsubsection{\textit{Actions}}
The set of actions (powers) $A$ is defined as $A=\left\lbrace a_1, a_2, ... , a_{N_{\text{power}}} \right\rbrace$, which uniformly covers the range between minimum ({\small$a_1=P_{min}$}) and maximum ({\small$a_{N_{\text{power}}}=P_{max}$}) power.


\subsubsection{\textit{States}}
We define $N_{r}$ equally spaced concentric circles around the cluster head (CH) of each cluster. These circles, define $N_r$ rings with $r$ units of spacing, around the CH. The state of the agent $k$ at time step $t$ is defined as $\mathbf{s}_t^k=\left(n\right)$ which shows the ring number that the agent belongs to. Considering the definition of the Q-table and the states at the beginning of this section, if the agents' location is fixed, each agent will choose just one row of its Q-table to search for the best action decision.

\subsubsection{\textit{Reward}}
$R_t^k$ is the immediate reward incurred due to selection of the action $a_t^k$ at state $\mathbf{s}_t^k$ by the agent $k$ at time step $t$. The constraint in~\eqref{c}, can be represented as: $C_t^k \geq \log_2(q_k) ,$ for $k = 1,...,N$. $C_t^k$ is the normalized capacity of agent $k$ at time step $t$. Based on this, the normalized proposed reward function for the agent $k$ at time step $t$ is defined as
{
\begin{flalign}\label{RF1}
  R_t^k =
    \underbrace{\frac{1}{2\log_2\left(q_k\right)}}_{(a)} \left( \underbrace{C_t^k}_{(b)} - \underbrace{\left| C_t^k-2\log_2\left(q_k\right)\right|}_{(c)} \right).
\end{flalign}
}
The rationale behind the proposed reward function is as follows

\begin{figure}[h]
\begin{centering}
\includegraphics[width=0.7\columnwidth]{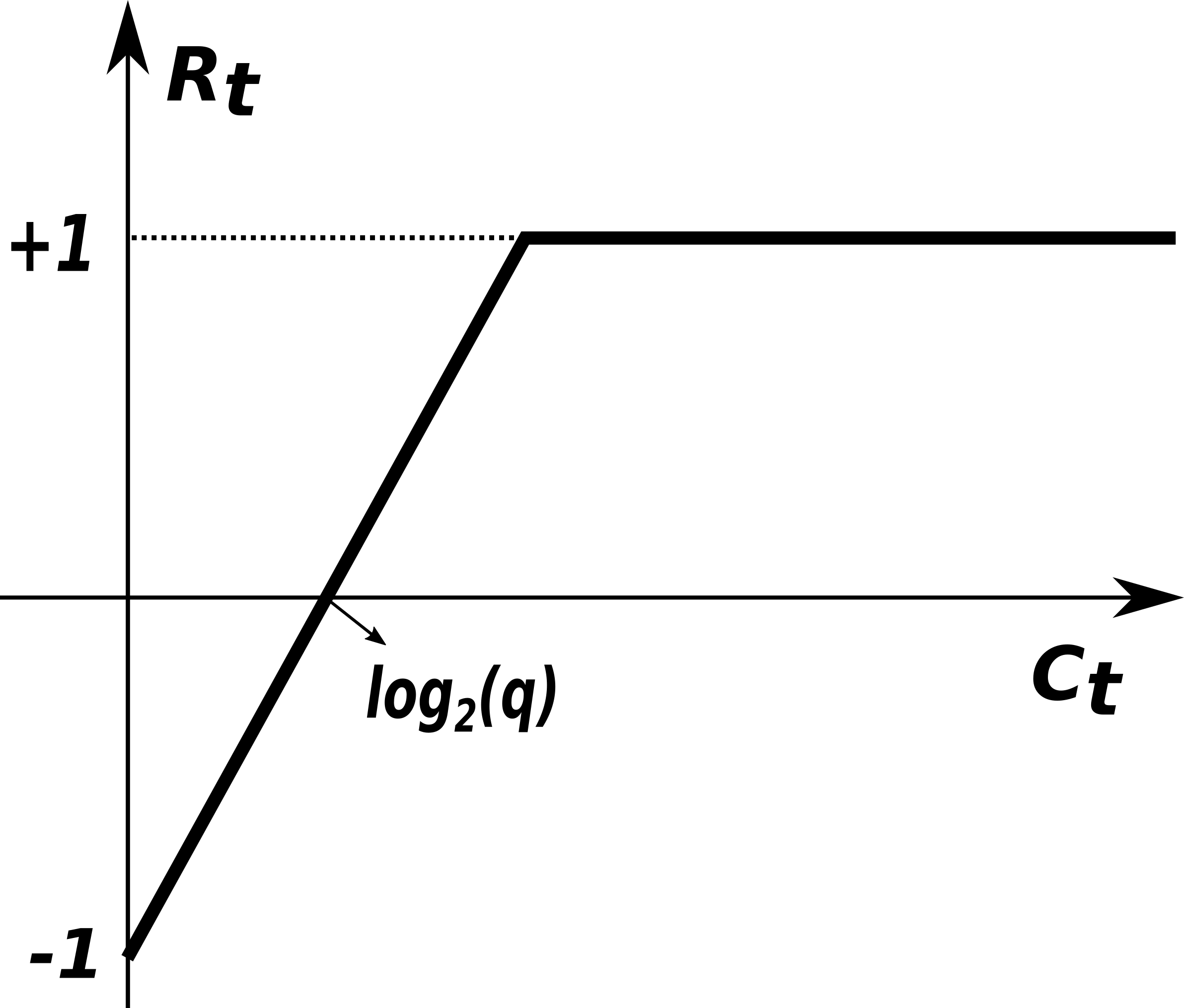}
\caption[width=.3\textwidth]{Proposed reward function (RF).}\label{fig_RF}
\end{centering}
\end{figure}

\begin{itemize}
\item The term (\textit{a}) normalizes the value of reward function.
\item The objective of the optimization problem is to maximize the capacity of the network, so the term (\textit{b}) results in a higher reward for higher capacity for an agent.
\item  To satisfy the QoS constraint for agent $k$, capacity deviation of its associated user from the required QoS, term (\textit{c}), should result in a lower reward.
\item There is a maximum reward ($+1$) for an agent to provide fairness between the agents which is shown in Fig.~\ref{fig_RF}.
\item The proposed reward function is a first order function of $C_t^k$, which reduces each iteration's complexity. 
\end{itemize}

\vspace{-10pt}
\section{Simulation Results} \label{sec_sim}
In this section the simulation setup is detailed and then the results of the simulations are presented.
\vspace{-10pt}
\subsection{Simulation Setup}\label{sec_setup}
A dense mmWave BS network, with approximately $120$ BSs in a $1~km^2$ area is considered. The BSs are distributed based on SPPP and operate independently in the network. Each BS, supports one user equipment (UE), which is located in a radius of $10$m around the BS. The QoS for a user is defined as the required $\text{SINR}$ to support the user's service. The value of $q_k=2.83$ is considered for all the users.

To perform Q-\textit{learning}, the learning rate is considered as $\alpha=0.5$, the discount factor as $\gamma=0.9$, $N_{\text{power}} = 31$, $r=50$m, and $N_r=4$. The maximum number of iterations is set to $50,000$. The remaining parameters of the simulation are represented in Table~\ref{table_params}.
\begin{table}[h]
\centering
\caption{Simulation Parameters}\label{table_params}
\begin{tabular}{|c|c|c|c|}
\hline
\textit{Param.} & \textit{Value} & \textit{Param.} & \textit{Value} \\ \hline
f               & 28 GHz         & $P_{min}$       & -10 dBm     \\ \hline
$\zeta$         & 8.7 dB         & $P_{max}$       &  35 dBm     \\ \hline
$\beta_1$       & 72.0           & $\beta_2$       & 2.92        \\ \hline
$\sigma^2$      & -120 dBm       &                 &              \\ \hline
\end{tabular}
\end{table}

\vspace{-10pt}

\subsection{Clustering Results}\label{sec_clusterResults}

\begin{figure}[h]
	\centering
	\begin{minipage}{0.7\columnwidth}
        \centering
        \includegraphics[width=1\columnwidth]{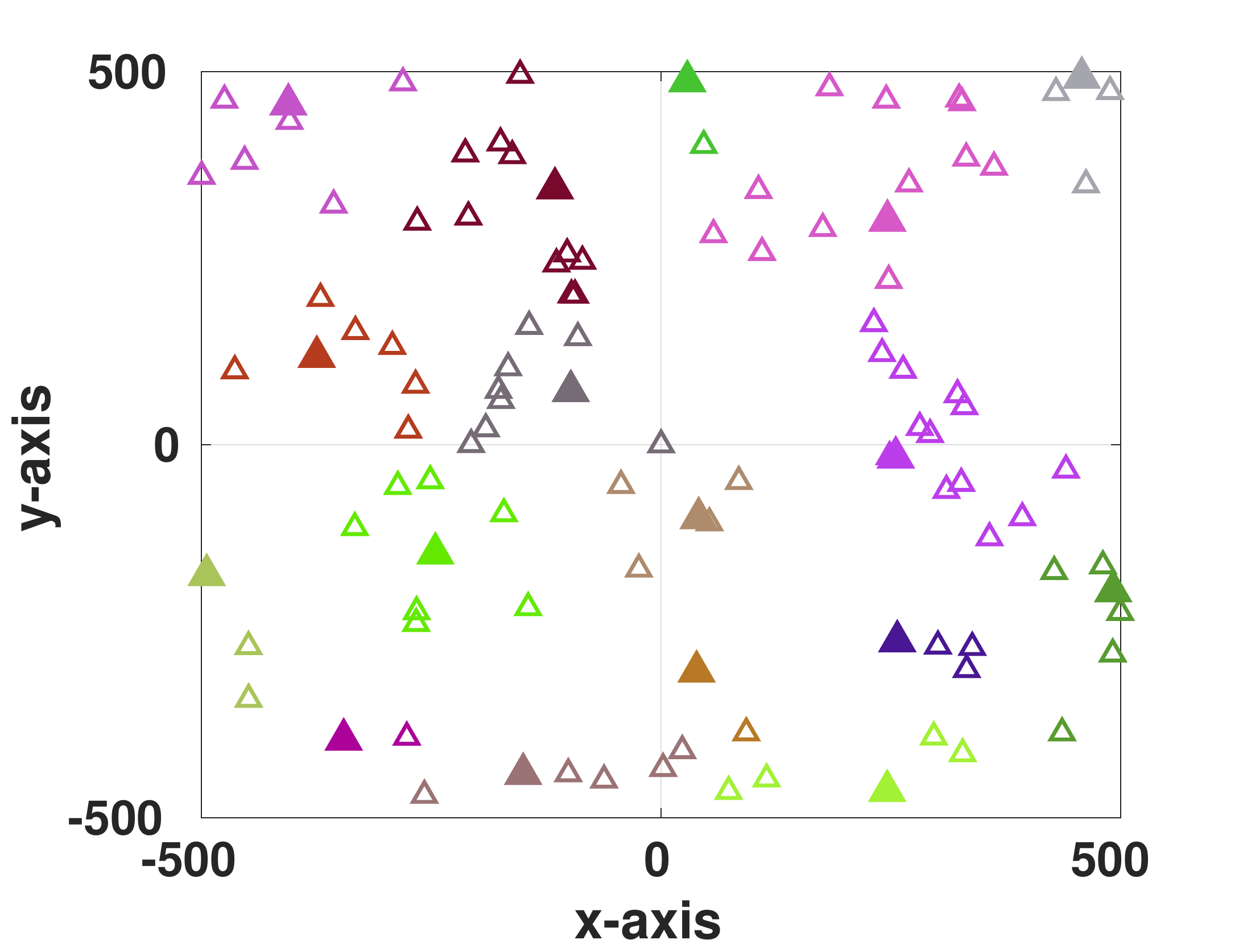} 
        \caption{\small $124$ BSs in $1~km^2$.}\label{fig_clusters1}
    \end{minipage}\hfill
    \begin{minipage}{0.7\columnwidth}
        \centering
        \includegraphics[width=1\columnwidth]{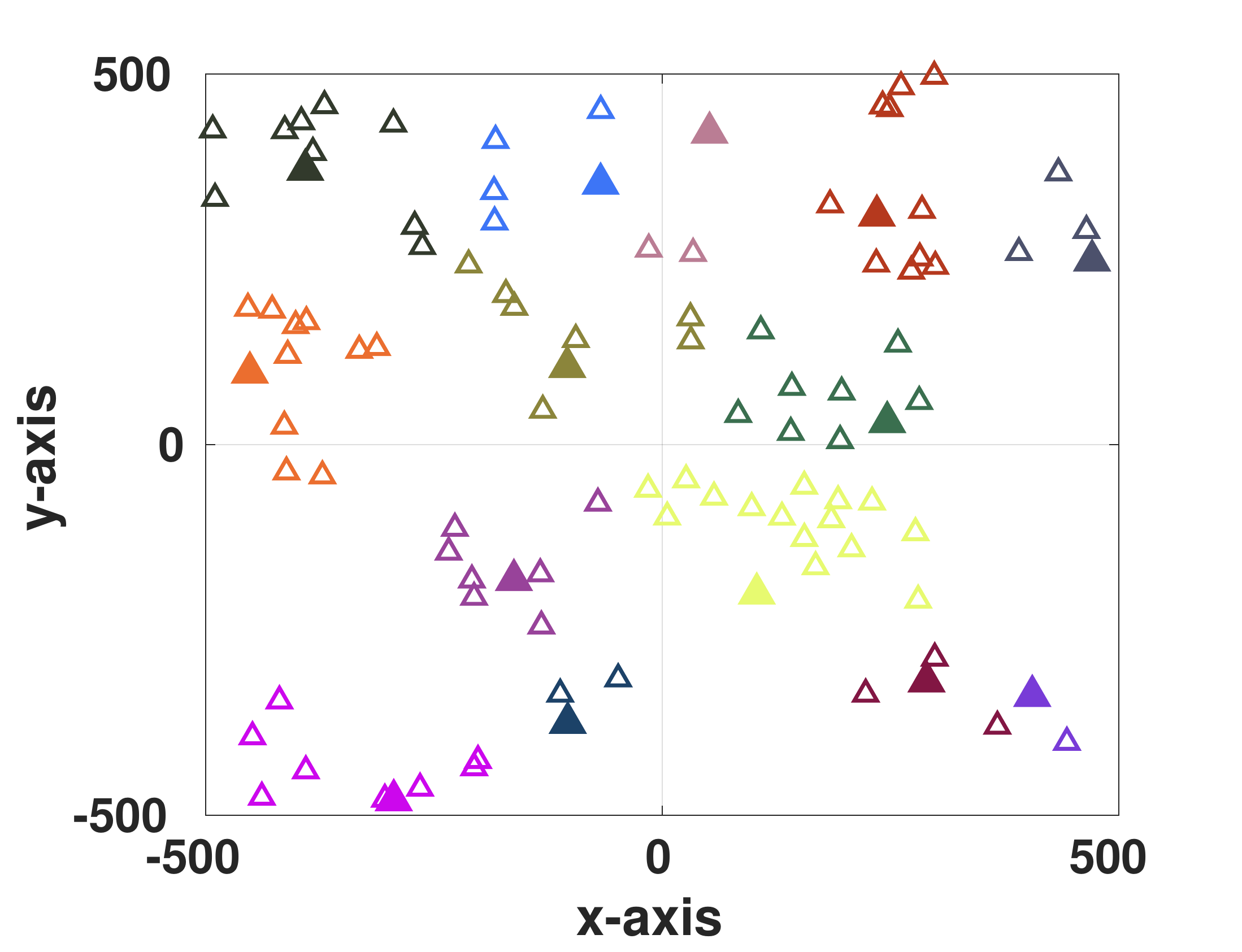} 
        \caption{\small $122$ BSs in $1~km^2$.}\label{fig_clusters2}
    \end{minipage}
\end{figure}

The implementation of clustering algorithm, is an event driven, message-passing distributed program in C++. Every BS is simulated as an independent thread, and is added to the network randomly in $[0,10]~\text{seconds}$. The clustering algorithm converges in less than $15~\text{seconds}$ for the assumed value for the $\lambda_{BS}$. The resulted clusters in two different distribution of BSs are shown in different colors in Fig.~\ref{fig_clusters1}, and~\ref{fig_clusters2}. Each cluster head (CH) is marked with a filled color.

\vspace{-10pt}
\subsection{Power Allocation Results}

According to~\cite{art_dense0, art_rappaport}, the coverage range of millimeter communication is in the range of $100-200$ m, which means the maximum coverage of $0.12~(km^2)$ for each mmWave BS. Considering the interference-limited assumption and the value of $\lambda_{BS}=120~(\text{BS}/\text{km}^2)$, a cluster might have $14$ mmWave BSs. Hence, the \textbf{CDP-Q} algorithm results in clusters that include 2 to 14 BSs. 

\begin{figure}[h]
    \centering
    \begin{minipage}{0.75\columnwidth}
        \centering
        \includegraphics[width=1\columnwidth]{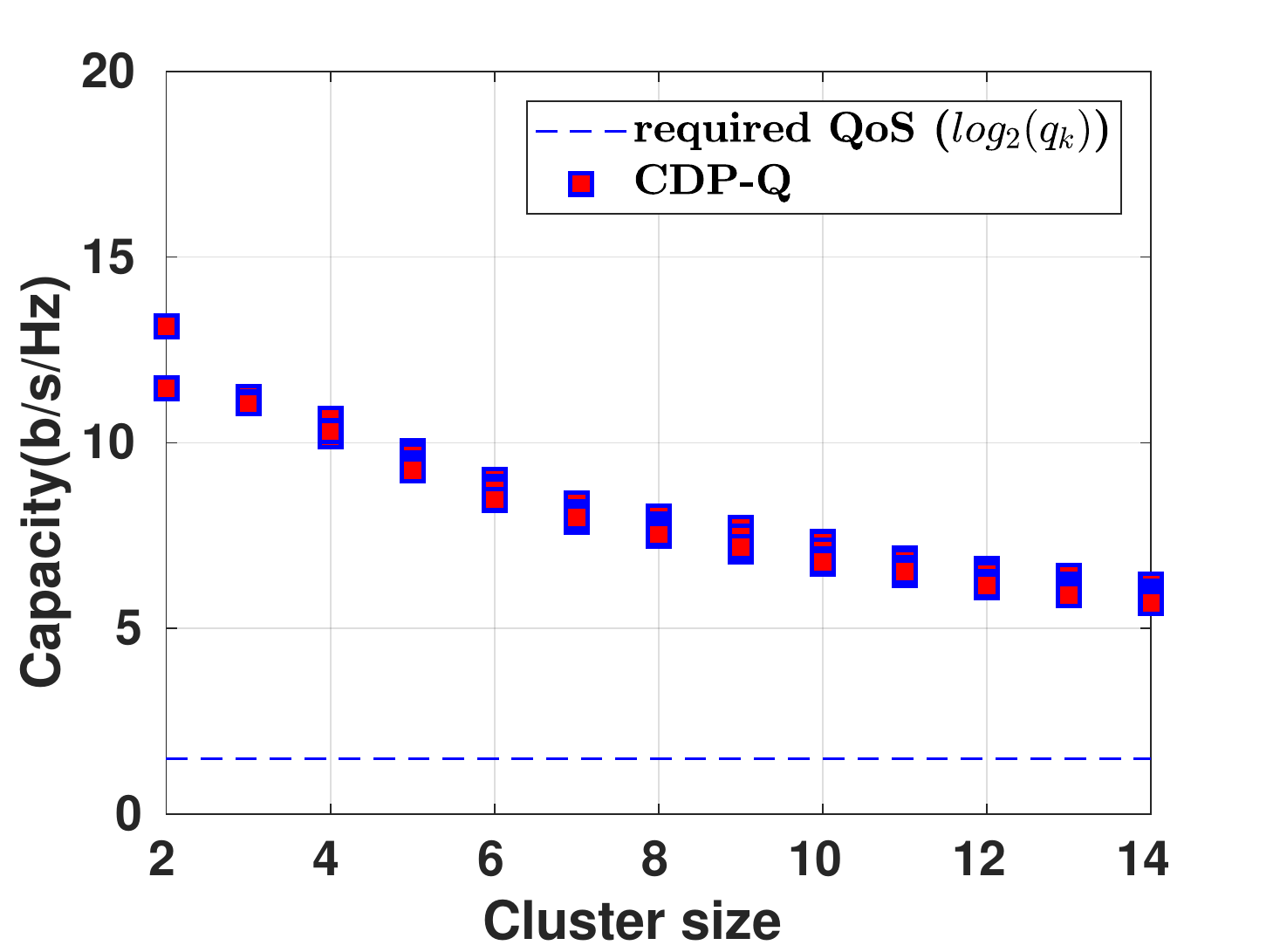}
		\caption{\small Capacity of clusters' members.}\label{fig_CDPQ}
    \end{minipage}\hfill
    \begin{minipage}{0.7\columnwidth}
        \centering
        \includegraphics[width=1\columnwidth]{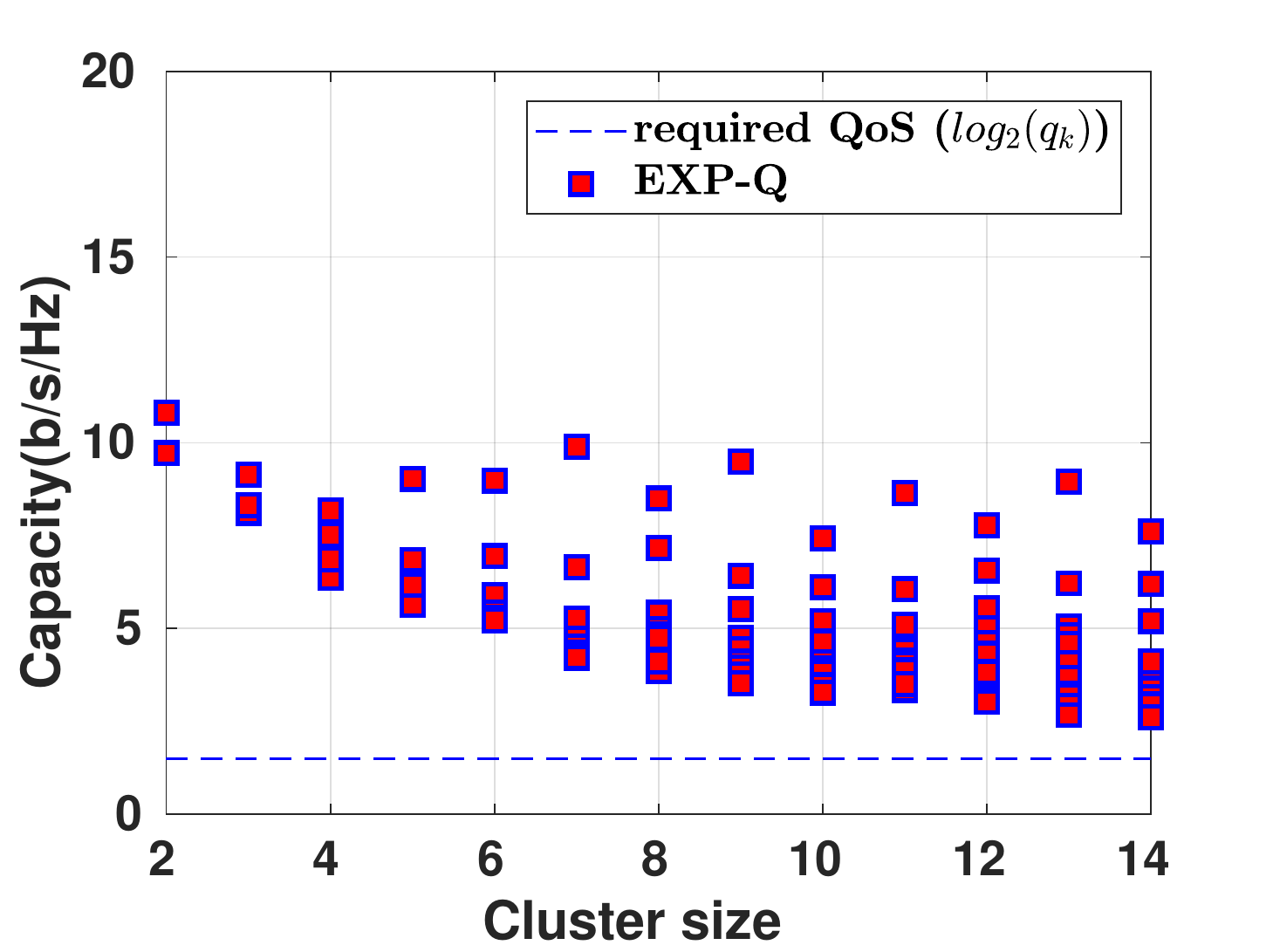} 
        \caption{\small Capacity of clusters' members.}\label{fig_EXPQ}
    \end{minipage}
\end{figure}

The results of power allocation using the proposed reward function are compared to the exponential reward function proposed in~\cite{art_reward2}, which are presented as EXP-Q in the simulations. For all possible cluster sizes, power allocation using the proposed reward function is simulated, and the normalized capacity of all BSs in the clusters are plotted in Fig.~\ref{fig_CDPQ}. The same simulations for EXP-Q are presented in Fig.~\ref{fig_EXPQ}.

\begin{figure}[h]
    \centering    
    \begin{minipage}{0.7\columnwidth}
        \centering
        \includegraphics[width=1\columnwidth]{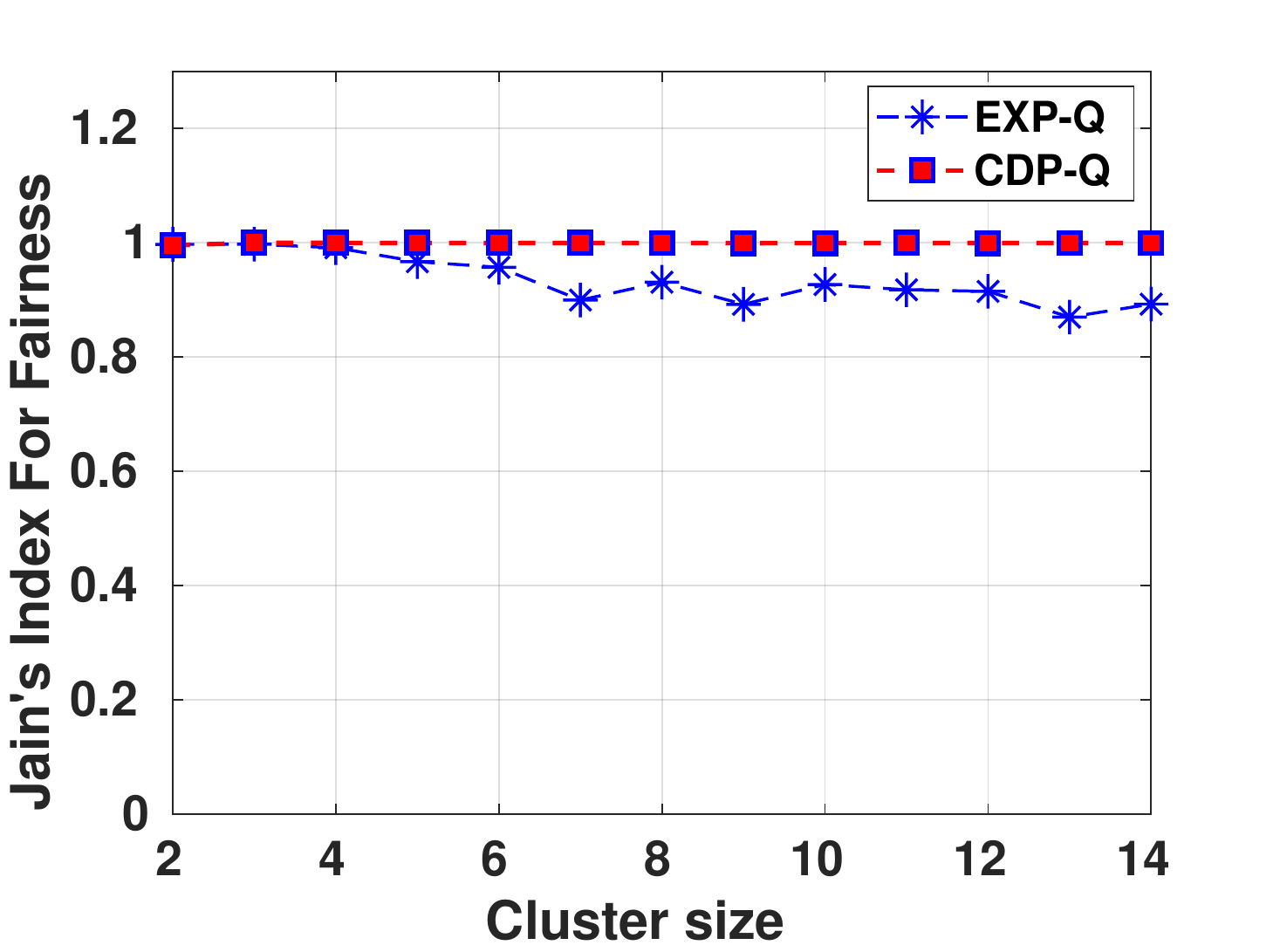}
		\caption{\small Jain's fairness index.}\label{fig_fair}
    \end{minipage}\hfill
    \begin{minipage}{0.7\columnwidth}
        \centering
        \includegraphics[width=1\columnwidth]{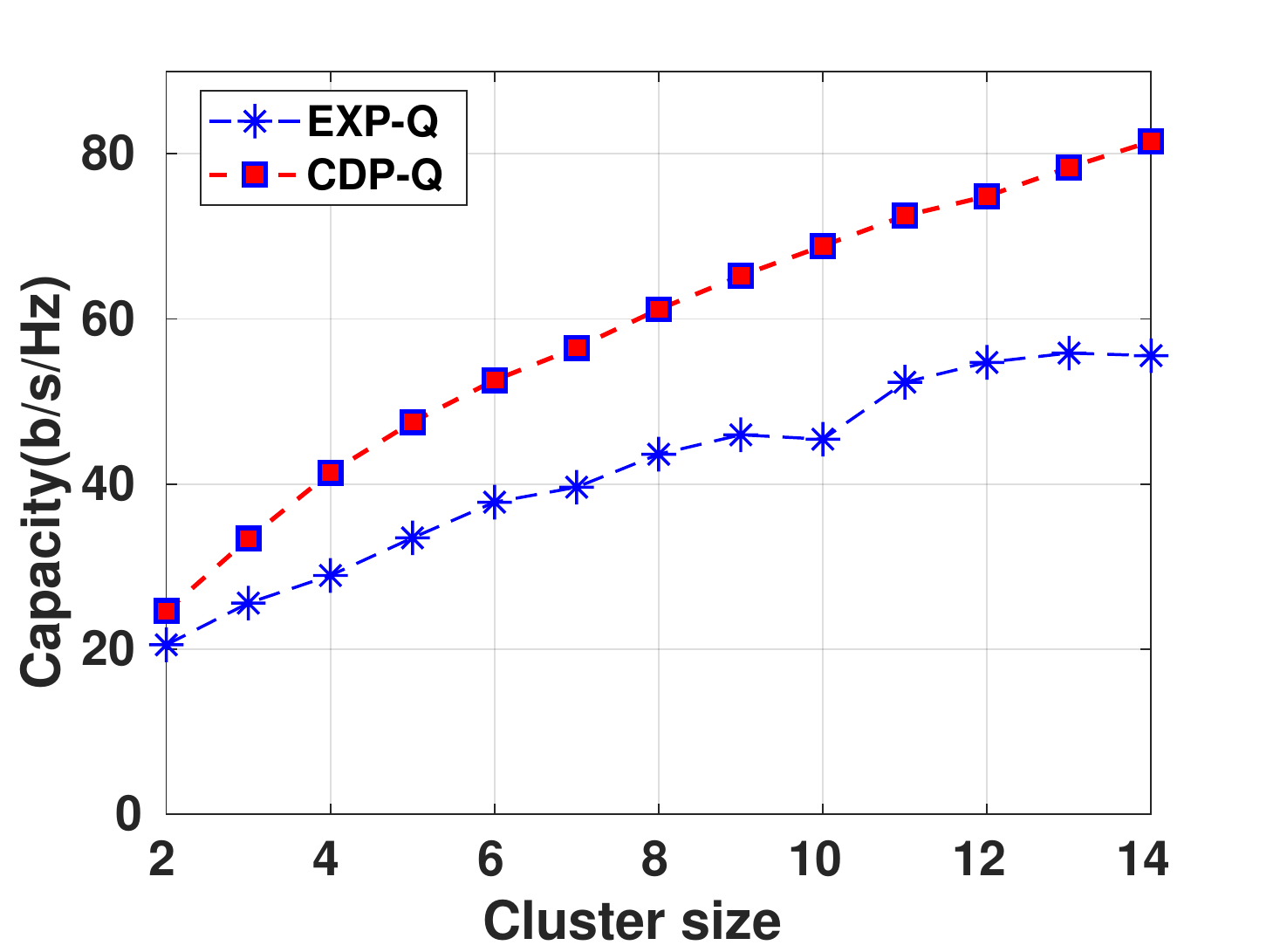} 
        \caption{\small Sum capacity of clusters.}\label{fig_totalC}
    \end{minipage}
\end{figure}

As it is shown in these figures, while both reward functions satisfy all members for all sizes of clusters with their required QoS, the normalized capacity of users in the CDP-Q are close to each other, while in EXP-Q the normalized capacity of users are much diverse. The diversity of normalized capacity values in EXP-Q effects the fairness index. The fairness index in each cluster is measured using Jain's fairness index~\cite{art_Jains} and is shown in Fig.~\ref{fig_fair}. In Fig.~\ref{fig_fair}, the CDP-Q maintains fairness for all sizes of clusters, while EXP-Q fails to support users with fairness in large cluster sizes. On the other hand, total capacity of the clusters are shown in Fig.~\ref{fig_totalC} with respect to the cluster size. According to Fig.~\ref{fig_totalC}, the CDP-Q provides higher capacity than the EXP-Q for all sizes of clusters. 

\section{Conclusion} \label{sec_con}
In this paper, a self-organized distributed power allocation algorithm is presented. The proposed algorithm reduces the optimization complexity by using a distributed clustering method, and provides adaptability in power allocation by using Q-\textit{learning}. The proposed reward function, satisfies required QoS for the users in all sizes of the resulted clusters, and outperforms the exponential based reward function.

\bibliographystyle{IEEEtran}
\bibliography{IEEEabrv,ref}

\end{document}